\documentclass[twocolumn,showpacs,preprintnumbers,amsmath,amssymb]{revtex4}

\usepackage{latexsym} 
\usepackage{graphicx}       
\usepackage{bm}   

\newcommand{\al}{\alpha}

\newcommand{\de}{\delta}
\newcommand{\De}{\Delta}
\newcommand{\ep}{\varepsilon}

\newcommand{\la}{\lambda}

\newcommand{\si}{\sigma}

\newcommand{\Om}{\Omega}

\newcommand{\beq}{\begin{equation}}
\newcommand{\eeq}{\end{equation}}
\newcommand{\ba}{\begin{array}}
\newcommand{\ea}{\end{array}}
\newcommand{\bea}{\begin{eqnarray}}
\newcommand{\eea}{\end{eqnarray}}
\newcommand{\bi}{\begin{itemize}}  
\newcommand{\ei}{\end{itemize}}
\newcommand{\ben}{\begin{enumerate}} 
\newcommand{\een}{\end{enumerate}}
\newcommand{\bc}{\begin{center}}
\newcommand{\ec}{\end{center}}

\newcommand{\p}{\partial}

\newcommand{\txt}{\textstyle}
\newcommand{\dsp}{\displaystyle}
\newcommand\eqn[1]{(\ref{#1})}      

\newcommand{\half} {{\txt \frac{1}{2}}}

\newenvironment{tightlist}[1]{ 
\begin{list}{#1}{
  \usecounter{enumi}
  \setlength{\topsep}{0ex} 
  \setlength{\itemsep}{-\parsep} 
  \settowidth{\labelwidth}{#1} 
  \setlength{\labelsep}{0.5em}        
  \setlength{\leftmargin}{\labelwidth}
  \addtolength{\leftmargin}{\labelsep}
 }}{\end{list}}

\newcommand{\fm}{{\rm fm}}

\newcommand{\MeV}{{\rm MeV}}

\newcommand{\chiQ}{\raisebox{0.2ex}{$\chi$}^{\phantom{y}}_Q} %

\newcommand{\nQ}{n^{\phantom{y}}_Q} 
\newcommand{\mucrit}{\mu_{\rm crit}}
\newcommand{\sicrit}{\si_{\rm crit}}
\newcommand{\pcrit}{p_{\rm crit}}
\newcommand{\Rcell}{{R_{\rm cell}}}
\newcommand{\Rstar}{{R_{\rm star}}}
\newcommand{\pext}{p_{\rm ext}}
\newcommand{\pQM}{p^{\phantom{y}}_{\rm QM}} 
\newcommand{\mue}{\mu_{e}} 
\newcommand{\sn}{{\rm sn}}

\usepackage[normalem]{ulem}  

\begin{document}

\title{Thickness of the strangelet-crystal crust of a strange star}
\author{Mark G. Alford and David A. Eby}
\affiliation{Physics Department, Washington University,
St.~Louis, MO~63130, USA}

\begin{abstract}
It has recently been pointed out that if the surface tension 
of quark matter is low enough,
the surface of a strange star will be a crust consisting of a
crystal of charged strangelets in a neutralizing background of electrons.
This affects the behavior of the surface, and must be taken into account
in efforts to observationally rule out strange stars.
We calculate the thickness of this ``mixed phase'' crust,
taking into account the effects of surface tension and Debye screening
of electric charge.
Our calculation uses a generic parametrization of the equation of
state of quark matter. 
For a reasonable range of quark matter equations of state,
and surface tension of order a few MeV/fm${}^2$,
we find that the preferred crystal 
structure always involves spherical strangelets, not rods or slabs
of quark matter.
We find that for a star of radius 10 km and mass $1.5 M_\odot$,
the strangelet-crystal crust can be from zero to
hundreds of meters thick,
the thickness being greater when the strange quark is heavier,
and the surface tension is smaller. For smaller quark stars
the crust will be even thicker.
\end{abstract}

\date{5 August 2008}


\maketitle

\section{Introduction}
\label{sec:intro}

Quarks in their most familiar form are confined in protons and
neutrons that make up standard nuclear matter. However, according to
the ``strange matter hypothesis'' 
\cite{Bodmer:1971we,Witten:1984rs}
this form of matter might be metastable and the fully 
stable state would then be ``strange matter'', 
which contains roughly equal numbers
of up, down, and strange quarks. Large (kilometer-sized)
pieces of strange matter are ``strange stars''
(for a review see Ref.~\cite{Weber:2004kj});
small nuggets of strange matter are 
``strangelets'' \cite{Farhi:1984qu}. 
The strange matter hypothesis remains a fascinating but unproven conjecture.
In this paper we will assume that it is correct, and investigate the
structure of the crust of a strange star.

The traditional picture of the surface of a strange star 
is a sharp interface, of thickness
$\sim$ 1 Fermi. Below the interface lies
quark matter, the top layer of which is positively charged. Above the
interface is a cloud of electrons, sustained by an electric
field which could also support a thin nuclear matter crust in
suspension above the quark matter~\cite{Alcock:1986hz,Stejner:2005mw}, as long
as the strange star is not too hot~\cite{Usov:1997eg}.

However, if the surface tension $\si$ of the interface
between quark matter and the vacuum is small enough, the surface will take on
a much more complicated structure. If $\si$ is less than a critical value
$\sicrit$ 
then large lumps of strange matter are unstable against fission into
smaller pieces \cite{Jaikumar:2005ne,Alford:2006bx}. As a result, the
simple surface described in the previous paragraph is unstable, and 
is replaced by a mixed phase
involving nuggets of positively-charged
strange matter in a neutralizing background of electrons. 
It is reasonable
to guess that the ground state is a regular lattice, leading to a
crust with a crystalline structure. (Note that this crystal is completely
different from the Larkin-Ovchinnikov-Fulde-Ferrell phase of quark
matter \cite{Alford:2000ze,Casalbuoni:2003wh}, 
where the quark matter density is uniform, but the pairing gap varies
in space.)
Jaikumar, Reddy, and Steiner
\cite{Jaikumar:2005ne} conjecture that the strangelet crystal
crust will actually be
a multi-layer structure of mixed phases, 
analogous to the ``nuclear pasta''
phases that occur in models of the inner crust of a conventional
neutron star \cite{Ravenhall:1983uh}.
At the outer edge of the crust, we expect a dilute low-pressure lattice of 
small strangelets in a degenerate gas of electrons. 
As we descend in to the star,
the pressure rises, and the structure is modified (becoming denser,
and perhaps changing to rods, slabs, cavities, etc). At a critical
pressure $\pcrit$ the mixed phase is no longer
stable, and there is a transition to uniform neutral quark matter.
As one burrows deeper into the star the pressure continues
to rise, and there may be other phase transitions between different
phases of quark matter \cite{Alford:2007xm}, but those will not concern us here.
Note that in this scenario the strange star depends on gravity for
its existence. In the absence of gravity, it would undergo fission into
strangelets.

Ref.~\cite{Jaikumar:2005ne}, assuming zero surface tension and
neglecting Debye screening, estimated that the mixed phase crust might
be $40-100$~m thick, with $\pcrit\approx 1000~\MeV^4$.
This is an interesting result because
if a strange star has a sufficiently thick crystalline crust, it
might be hard to distinguish from the crust of a neutron star.
Astrophysical properties that are sensitive to the crust include
cooling behavior, neutrino and photon opacity during a supernova,
the photon emission spectrum, glitches, and
frequencies of seismic vibrations which are observed after giant flares
in magnetars. For further discussion and references
see Sec.~\ref{sec:discussion}.
This paper will make a more careful calculation of the properties
of a strangelet crystal crust,
including the effects of Debye screening \cite{Heiselberg:1993dc}
and surface tension.

We expect that the properties of the crust will emerge from a
competition between various different contributions to the energy.
Charge separation is often favored by the internal energy
of the phases involved, because a neutral phase is always
a maximum of the free energy with respect to the electrostatic
potential (see \cite{Ravenhall:1983uh,Glendenning:1992vb};
for a pedagogical discussion see \cite{Alford:2004hz}).
The domain structure is determined by
surface tension (which favors large domains) and electric field
energy (which favors small domains). Debye screening is important
because it redistributes the electric charge, concentrating it
in the outer part of the quark matter domains and the inner part
of the surrounding vacuum, and thereby modifying the internal energy
and electrostatic energy contributions.

To make an estimate
of the thickness of the crust we need to 
calculate the equation of state of the mixed phase,
i.e.~the energy density $\ep_{\rm mp}$ as a function of 
the pressure $p_{\rm mp}$.
The thickness of the crust for a star of mass $M$ is then 
\beq
\Delta R = \Rstar\Bigl(\frac{\Rstar}{GM}-2\Bigr)\int_0^{\pcrit} 
  \frac{1}{\ep_{\rm mp}} dp_{\rm mp} \ ,
\label{dr}
\eeq
in $\hbar=c=1$ units. This expression follows from the 
Tolman Oppenheimer Volkoff equation \cite{Tolman:1939jz,Oppenheimer:1939ne},
assuming that $\De R\ll \Rstar$, and that everywhere
in the crust the pressure is much smaller than both the local energy density and
the average energy density of the whole star. These are very
good approximations for the cases that we study.

We obtain $\ep_{\rm mp}$ as a function of $p_{\rm mp}$
by dividing the strangelet lattice into unit cells
(``Wigner-Seitz cells'') and calculating the pressure at the edge of a
cell as a function of its energy density.  We study cells that are
three-dimensional (a lattice of strangelets in a degenerate gas of
electrons), two-dimensional (rods of strange matter in a degenerate
gas of electrons) and one-dimensional (slabs of strange matter
interleaved with regions of degenerate electron gas).
Our approach is similar to that used in studying mixed phases
of quark matter and nuclear matter in the interior of neutron stars
\cite{Maruyama:2007ey}.

We build on the formalism for a generic quark matter equation of state
and infinitely-large Wigner-Seitz cells
that was developed in \cite{Jaikumar:2005ne,Alford:2006bx}. 
The main assumptions that we make are:\\
1) Within each Wigner-Seitz cell we use a Thomas-Fermi approach,
solving the Poisson equation to obtain the charge distribution,
energy density, and pressure. This is incorrect for
very small strangelets, where the energy level structure
of the quarks becomes important \cite{Madsen:1994vp,Amore:2001uf}; such
corrections may be relevant for the very low pressure (outer crust)
part of our results (see Sec.~\ref{sec:discussion}).
\\
2) We assume our $D$-dimensional Wigner-Seitz cells to be $D$-spheres.
In reality the cells will be unit cells of some
regular lattice (cubic, hexagonal close packed, etc).
However, as long
as the cell is much bigger than the strangelet inside it, we expect this
approximation to be reasonably accurate. We will only report results for
cases where $\Rcell>2R$ ($R$ being the radius of the quark matter
in the center of the cell, which we expect will have a
rotationally symmetric shape because of the surface tension).
In some cases this assumption is
violated, and we will then only be able to
obtain a lower limit on the crust thickness 
(see Sec.~\ref{sec:thickness-results}).
\\
3) We treat the interface between quark matter and the vacuum as a
sharp interface, with no charge localized on it, which is characterized
by a surface tension. We neglect any surface charge that might arise
from the reduction of the density
of states of strange quarks at the surface 
\cite{Madsen:2000kb,Madsen:2001fu,Madsen:2008bx,Oertel:2008wr}.
We also neglect the curvature energy of a
quark matter surface \cite{Christiansen:1997rc,Christiansen:1997vt},
so we do not allow for ``Swiss-cheese'' mixed phases, in which 
the outer part of the Wigner-Seitz cell is filled with quark matter, with
a cylindrical or spherical cavity in the center, for which the curvature
energy is crucial. Note that these phases would be
expected to occur at higher pressure than the ones we study, so
including them is likely to make the crust even thicker than we predict. \\
4) We assume that the chemical potential for negative electric charge $\mue$ 
is much less than the chemical potential for quark number $\mu$. This allows us
to expand the quark matter equation of state in powers of $\mue$, and means that
within the quark matter we can ignore the contribution of electrons
to the charge and pressure. This is a very good approximation for small 
strange quark mass, which corresponds to
small $\nQ$ in our parameterization. For the largest value of $\nQ$
that we study, $\mue$ in neutral quark matter 
is close to 100 MeV, and the assumption is still reasonable.\\
5) We assume that only electrons are present, with no muons. This is valid
as long as $\mue$ is less than the muon mass $m_\mu$, which is true for
all the cases that we study.\\
6) We assume that $\mue$ is always much greater than the electron mass.
Thus in the degenerate electron gas, we can take the electrons to be massless,
which simplifies the Thomas-Fermi calculation of their charge distribution.
Since $\mue$ drops monotonically from the center of the cell to its edge,
this condition will only be violated for very large cells 
(very low pressures).\\
8) We always work at zero temperature. The temperature of the surface of
a compact star, even during a flare \cite{Lyubarsky:2002cs}, 
is expected to be less than 100 keV,
so we expect this to be a reasonable approximation.

\section{Characterization of quark matter}
\label{sec:characterization}

\subsection{Generic parametrization}
We use the generic parametrization of the quark matter equation of
state suggested in Ref.~\cite{Alford:2006bx},
\beq
\pQM(\mu,\mue)  =
 p_0(\mu)-\nQ(\mu)\mue + \half\chiQ(\mu) \mue^2+\ldots
\label{eqn:generic_EoS}
\eeq
which expresses the pressure as a function of the
chemical potential for quark number ($\mu$) and 
for negative electric charge 
($\mue$), expanded to second order in $\mue$. 
In addition, we assume that there is a surface tension $\si$
associated with the interface between quark matter and vacuum.
In this paper we do not include curvature energy.

This parametrization is model-independent.
Any specific model of quark matter can be represented by appropriate
choices of $\si$, $p_0$, the charge density $\nQ$, and charge susceptibility
$\chiQ$.

The quark density $n$ and the electric charge density 
$q^{\phantom{y}}_{\rm QM}$ 
(in units of the positron charge) are
\beq
n = \frac{\p \pQM}{\p \mu},\qquad
q^{\phantom{y}}_{\rm QM} =  -\frac{\p \pQM}{\p \mue}
 = \nQ - \chiQ\mue \ .
\label{charges}
\eeq
So in uniform neutral quark matter the electron chemical
potential is $\mue^{\rm neutral}=\nQ/\chiQ$. Eq.~\eqn{eqn:generic_EoS}
is a generic parametrization if
$\mue^{\rm neutral}\ll\mu$, which is typically the case in three-flavor
quark matter.

The bag constant enters in $p_0(\mu)$, and we will fix it by
requiring that the first-order 
transition between neutral quark matter and the vacuum
occur at quark chemical potential $\mucrit$, 
i.e.~$p(\mucrit,\mue^{\rm neutral})=0$.
Because we
are assuming that the strange matter hypothesis is valid, we must have
$\mucrit \lesssim 310$~MeV, since at $\mu\approx 310$~MeV there is
a transition from vacuum to neutral nuclear matter.
In this paper we will typically use $\mucrit=300~\MeV$. 
The value of $\mu$ inside our quark matter lumps will always be very close
to $\mucrit$, so we can evaluate $\nQ$ and $\chiQ$ at $\mucrit$, and not
be concerned about their $\mu$-dependence.

We will restrict ourselves
to values of the surface tension that are below the critical value
\cite{Alford:2006bx} 
\beq
\ba{rcl}
\sicrit &=&\dsp 0.1325 \,
\frac{n_Q^2 \lambda_D}{\chi_Q}
=0.1325\,\frac{n_Q^2}{\sqrt{4\pi\alpha}\chi_Q^{3/2}},
\ea
\label{sicrit_result}
\eeq
where $\alpha=1/137$ and $\lambda_D$ is the Debye length 
\beq
\la_D = \frac{1}{\sqrt{4\pi\alpha\chi_Q}} \ .
\label{Debye_length} 
\eeq 
For typical models of quark matter, $\sicrit$ is of order $1$ to 
$10~\MeV/\fm^2$ (see Table~\ref{tab:crusts}).
If the surface tension is larger than $\sicrit$
then the energetically favored structure
for the crust will not be a strangelet crystal but
the simple sharp surface that has has been
assumed in the past \cite{Alcock:1986hz,Usov:1997eg}.

\subsection{Specific equations of state}
\label{sec:specific}

When we show numerical results we will need to vary $\nQ$ and $\chiQ$
over a range of physically reasonable values. To give a rough idea
of what values are appropriate, we consider the example of
non-interacting three-flavor quark matter, for which $\nQ$ and $\chiQ$
become functions of $\mu$ and the strange quark mass
$m_s$, while $p_0$ is in addition
a function of the bag constant $B$. Expanding to lowest non-trivial
order in $m_s$,
\beq
\ba{rcl}
p_0(\mu) &=& \dsp \frac{9\mu^4}{12\pi^2} -B  \ ,\\[2ex]
\nQ(\mu,m_s) &=& \dsp \frac{m_s^2\mu}{2\pi^2} \ ,\\[2ex]
\chiQ(\mu,m_s) &=& \dsp \frac{2\mu^2}{\pi^2}.
\label{unpaired}
\ea
\eeq
We emphasize that these expressions are simply meant to give a rough
idea of reasonable physical values for $\nQ$ and $\chiQ$. Our treatment
does not depend on an expansion in powers of $m_s$.
To tune the transition between neutral quark matter and the vacuum
so it occurs at $\mu=\mucrit$ (see previous subsection), we set $B$ so that
$p_0(\mucrit)=\half n_Q^2(\mucrit)/\chiQ(\mucrit)$.

In the regions between lumps of strange matter, we will assume that
there is a degenerate  electron gas, which we treat in the Thomas-Fermi approximation.
As long as $\mue$ is much greater than the electron mass, we can treat
the electrons as massless particles, whose pressure and charge density
(in units of $e$) is
\beq
p_{e^{\!-}}(\mue) = \dsp \frac{\mue^4}{12\pi^2} \ , \qquad
q_{e^{\!-}}(\mue) = \dsp -\frac{\mue^3}{3\pi^2} \ .
\label{electrons}
\eeq

\section{Analysis of a Wigner-Seitz cell}
\label{wigner-seitz}

We will study one, two, and three dimensional Wigner-Seitz cells. In the center
there is a slab, rod, or sphere of quark matter, with radius $R$. 
The cell itself has radius
$\Rcell$. We want to calculate the equation of state of a mixed phase
made of Wigner-Seitz cells, so we 
solve for the charge density, energy density, and
pressure throughout the cell, using the Thomas-Fermi approximation
for the contributions of quarks and electrons.
This corresponds to solving the
the Poisson equation, which reads (in Heaviside-Lorentz 
units with $\hbar=c=1$)
\beq
\nabla^2 \mue(r) = -4 \pi \alpha q(r) \ ,
\label{Poisson}
\eeq
where $q(r)$ is the electric charge density in units of the positron charge $e$,
and $\mue$ is the electrostatic potential divided by $e$. The
equation is not trivial to solve because the charge density is 
itself a function
of $\mue$ (see \eqn{charges}).
The boundary conditions are that there is no electric field in the
center of the cell (no $\de$-function charge there), and
no electric field at the edge of the cell (the cell is electrically neutral),
\beq
\frac{d \mue}{d r}(0) = 0 \ , \qquad
\frac{d \mue}{d r}(R_{\rm cell}) = 0 \ . 
\label{BC}
\eeq
We also need a matching condition at the edge of the quark matter,
i.e.~at $r=R$. As discussed in Sec.~\ref{sec:intro},
we assume that no charge localized on the surface, so we
require continuity of the potential and electric field at $r=R$,
\beq
\mue(R\!+\!\de) = \mue(R\!-\!\de) \ , \qquad
\frac{d \mue}{d r}(R\!+\!\de) = \frac{d \mue}{d r}(R\!-\!\de) \ .
\label{MC}
\eeq

In two or three dimensional cells, the
value of $\mu$ inside the strange matter will be slightly
different from $\mucrit$ because the surface
tension compresses the droplet.
To determine the value of $\mu$, we require the pressure discontinuity
across the surface of the strangelet to be balanced by the surface tension:
\beq
\pQM(\mu,\mue(R)) -  p_{e^{\!-}}(\mue(R)) = \frac{(D-1)\si}{R} \ .
\label{pressure_disc}
\eeq

Once these equations are solved, we can obtain the relevant properties
of the cell. The total energy of a $D$-dimensional Wigner-Seitz cell is
\beq
\ba{rcl}
E &= & \dsp \int_0^R \! \Om_D(r)dr \, \Bigl(
  \mu n(\mue) - \half \mue q_{\rm QM}(\mue) - \pQM(\mu,\mue)\Bigr)  \\[3ex]
 &+& \dsp \int_R^\Rcell \!\Om_D(r)dr \,
    \Bigl( - \half \mue q_{e^{\!-}}(\mue) - p_{e^{\!-}}(\mue) \Bigr) \\[3ex]
 &+& \dsp \Om_D(R)\, \si \ ,
\ea
\label{energy}
\eeq
where $\mue$ is a function of $r$, and
$\Om_D(r)$ is the surface areas of a $(D-1)$-sphere, i.e.
\beq
\Om_1 = 2\ , \qquad \Om_2(r)=2\pi r \ ,\qquad \Om_3(r)=4\pi r^2 \ .
\label{measure}
\eeq
The $-\half \mue q$ terms in \eqn{energy} 
come from combining $-\mue q$ (from the relationship
between energy density and pressure) with the electric field energy density
$+\half \mue q$.
The external pressure of the cell is simply the pressure of the electrons
at the edge of the cell,
\beq
\pext = p_{e^{\!-}}(\mue(\Rcell)) \ .
\label{pext}
\eeq
The total number of quarks is
\beq
N = \int_0^R \! \Om_D(r)dr \, n(\mu,\mue) \ .
\eeq
The volume of the cell is $V= 2 \Rcell$, $\pi\Rcell^2$, or $(4/3)\pi\Rcell^3$
for $D=1,2,3$ respectively, because we are assuming
rotationally symmetric cells.

By varying $R$ and $\Rcell$ we generate a two-parameter family of
strangelets. However, there is really only a single-parameter
family of physical configurations, parameterized by the external
pressure $\pext$. On each line of constant $\pext$ in the
$(R,\Rcell)$ parameter space, we must minimize the enthalpy per quark,
\beq
h = \frac{E + \pext V}{N} \ ,
\label{eq:enthalpy}
\eeq
to find the favored configuration.
We are at zero temperature so $h$ is also the
Gibbs free energy per quark.
This is done separately for $D=1,2,3$ cells, and the structure with the
lowest $h$ is the favored one.

We now have a well-defined way to obtain the equation of state
of the mixed phase of quark matter, namely
the energy density $\ep_{\rm mp}=E/V$ as a function
of the pressure $p_{\rm mp}=\pext$. This, via \eqn{dr},
determines the thickness of the strangelet crust.

\section{Solutions for the Wigner-Seitz cell}
\label{sec:solutions}

\subsection{Quark matter}
\label{quark-matter}
Inside the quark matter, we can solve the Poisson equation \eqn{Poisson} 
analytically. We can rewrite it
using \eqn{eqn:generic_EoS} and \eqn{charges} as
\beq
\nabla^2 \mue(r) = -4\pi\alpha(n_Q-\chi_Q \mue(r)) .
\label{eqn:Poisson2}
\eeq
In $D=1$, 2, or 3, the solutions
obeying the first boundary condition in \eqn{BC} are
\beq
\ba{rcl}
\mu_{e,1D}(r)&=&\dsp\frac{n_Q}{\chi_Q}+A \cosh(\frac{r}{\lambda_D}) \ ,\\
\mu_{e,2D}(r)&=&\dsp\frac{n_Q}{\chi_Q}+A J_0(\frac{ir}{\lambda_D}) \ ,\\
\mu_{e,3D}(r)&=&\dsp\frac{n_Q}{\chi_Q}+
\frac{A}{\lambda_D r} \sinh(\frac{r}{\lambda_D}) \ .
\ea
\label{mue-qm}
\eeq
The function $J_0$ is the zeroth order Bessel function of the first kind,
and it is a function of the square of its argument, so the result
is always real. 
The integration constant $A$ 
will be determined by matching to the solution outside the strange matter.

\subsection{Electron gas}
In the degenerate electron gas region outside the strange matter, 
from \eqn{electrons} and \eqn{Poisson} the
Poisson equation becomes 
\beq
\nabla^2 \mue(r) = \frac{4\alpha}{3\pi} \mue(r)^3  .
\label{eqn:Poisson-vac}
\eeq
There are three ways in which we were able to solve this equation.
In $D=1$, there is an exact analytic solution, which we present below.
In any number of dimensions there is an approximate
analytic solution, obtained by perturbing in powers of $\al$,
which works as long as the cell is not too large.
Finally, one can use brute-force
numerical methods to solve the differential equations with
the appropriate boundary conditions.
We used all three methods, checking their agreement with each other
in situations where more than one of them was applicable.
In our numerical results we give the values obtained by
numerical solution of the Poisson equation.

\subsubsection{Analytic solution for slabs}
\label{ssec:Complete}

In one dimension, the Poisson equation is
\beq
\frac{d^2\mue}{dr^2} = \frac{4\al}{3\pi}\mue^3 \ .
\eeq
By a change of variable to $\varphi = i\sqrt{2\al/3\pi}\mu_e$,
this becomes
\beq
\frac{d^2\varphi}{dr^2} = -2\varphi^3 \ ,
\label{vacD1}
\eeq
which belongs to a class of differential equations whose solutions
are Jacobi elliptic functions
$\sn(r|m)$ where $m$ is the ``parameter'' \cite{Weisstein}.
(Some authors write this as 
$\sn(r,k)$ where $m=k^2$ and $k$ is the elliptic modulus.)
The Jacobi elliptic function obeys
\beq
\frac{d^2\sn(r|m)}{dr^2} = -(1+m)\,\sn(r|m)+2m\, \sn(r|m)^3 \ ,
\eeq
which reduces to \eqn{vacD1} for $m=-1$,
so the closed form solution is
\beq
\varphi(r) = \frac{1}{X}\sn\Bigl(\frac{r-\Rcell}{X}+iK(-1) \ ,
  \,\Bigl|\Bigr.-1\Bigr)
\eeq
where $X$ is an integration constant, to be fixed by matching to
the quark matter solution at $r=R$. The argument is shifted by $iK(-1)$
to ensure that $\varphi'(\Rcell)=0$ 
(the boundary condition \eqn{BC} at $r=\Rcell$).
$K(m)$ is the complete elliptic integral of the first kind, 
so $K(-1)\approx 1.3110288$. Then $\varphi$ is purely imaginary,
and undoing the change of variables, the solution for 
$\mu_e$ is
\beq
\mu_{e,1D}(r) = -i\sqrt{\frac{3\pi}{2\al}}\frac{1}{X}\,
  \sn\Bigl( \frac{r-\Rcell}{X}+iK(-1)\Bigl|\Bigr. -1 \Bigr)\, .
\label{eqn:Jacobi}
\eeq
For one-dimensional slabs of quark matter we now have a complete
analytic solution, combining \eqn{mue-qm} at $r<R$ with \eqn{eqn:Jacobi}
at $r>R$, with $X$ and $A$ fixed by \eqn{MC}.

\subsubsection{Perturbative solution for cylinders and spheres}
\label{ssec:Analytic}

Another approach to solving \eqn{eqn:Poisson-vac},
which works in any number of dimensions,
is to expand in powers of
the electromagnetic coupling strength $\al$.
We write
\beq
\mue(r) = \mu_0(r)+\alpha \mu_1(r) +\alpha^2 \mu_2(r) +\cdots \ .
\eeq
Substituting in to  \eqn{eqn:Poisson-vac} and identifying
powers of $\al$, we find that to order $\al$,
\beq
\nabla^2 \mu_0(r) = 0 \ ,\qquad
\nabla^2 \mu_1(r) =  \frac{4}{3\pi} \mu_0(r)^3 
\label{eqn:Pert2}
\eeq
Solving these equations in $D=2,3$, we find
\beq
\ba{r@{}l}
%
\mu_{e,2D}&(r) = \dsp B+C \log(r/\lambda_D)
  +\frac{\alpha r^2}{6\pi}\Bigl[ 2B^3-6B^2 C \\[1ex]
&\dsp +\  9B C^2 -6C^3)+3(2B^2-4B C+3C^2)\log(r/\lambda_D)\\[1ex]
&\dsp +\ 6(B-C)C^2\log(r/\lambda_D)^2 
  + 2 C^3 \log(r/\lambda_D)^3\Bigr]\ ,\\[2ex]
\mu_{e,3D}&(r)= \dsp B-\frac{C}{r}+
\frac{2\alpha}{9\pi r}\Bigl[B^2 r^2(9C + B r) \\[1ex]
&\dsp -\  6C^2(C-Br)\log(r/\lambda_D)\Bigr] \ .
\ea
\label{pertsol}
\eeq
In each equation, the first two terms on the right-hand side
are the vacuum solution $\mu_0$, and the remainder are
the first-order correction.
The integration constants $B$ and $C$, along with
$A$ from \eqn{mue-qm}, are determined by the boundary 
condition \eqn{BC} at $r=\Rcell$ and the matching condition \eqn{MC}.

The perturbative solution works when screening is a small correction
to the unscreened (zeroth-order) electrostatic potential. It is
most reliable in three dimensions,
where the zeroth-order electrostatic potential becomes small at large $r$.
We used it to check our numerical results.

\begin{figure*}
\includegraphics[scale=0.23]{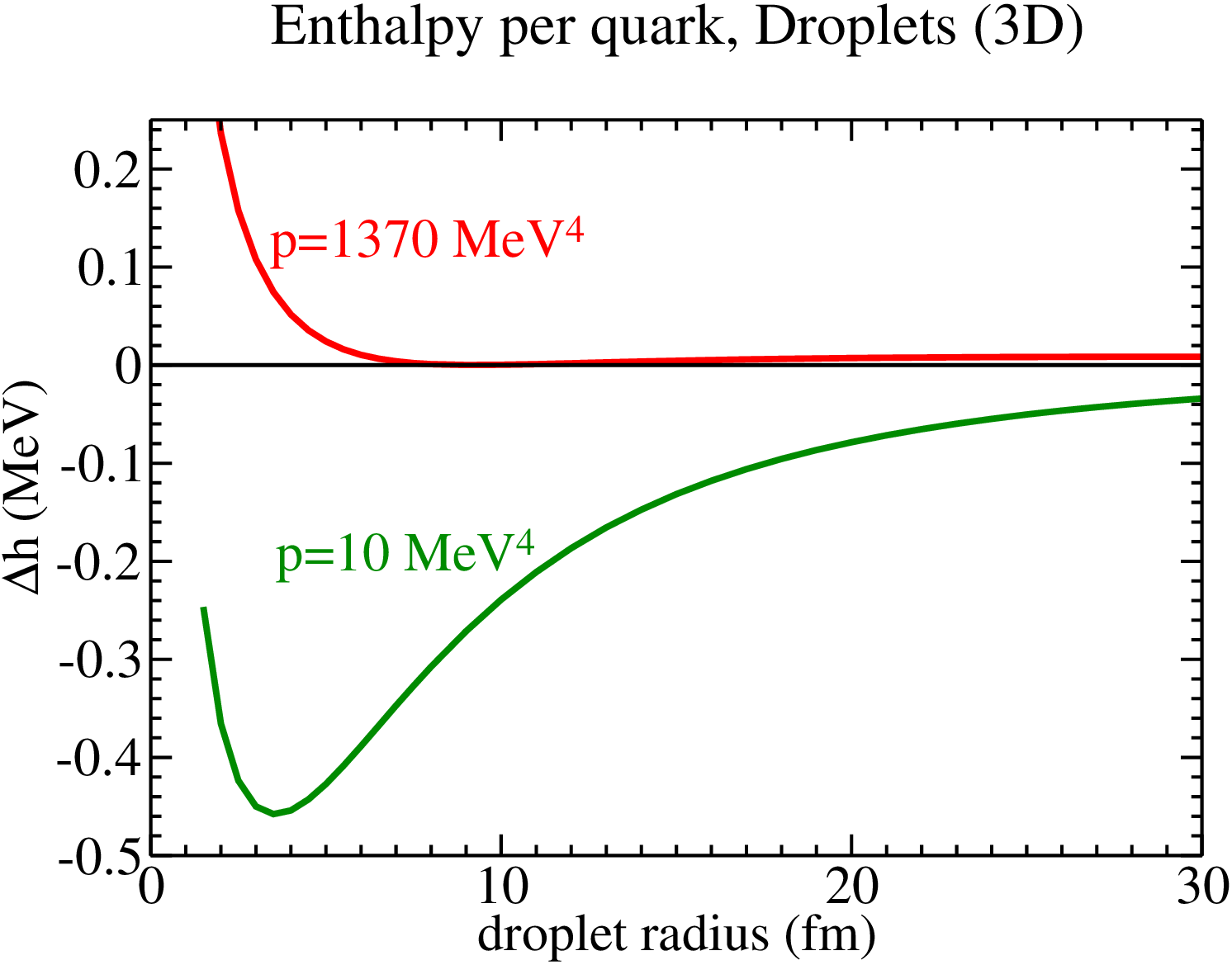}
\includegraphics[scale=0.23]{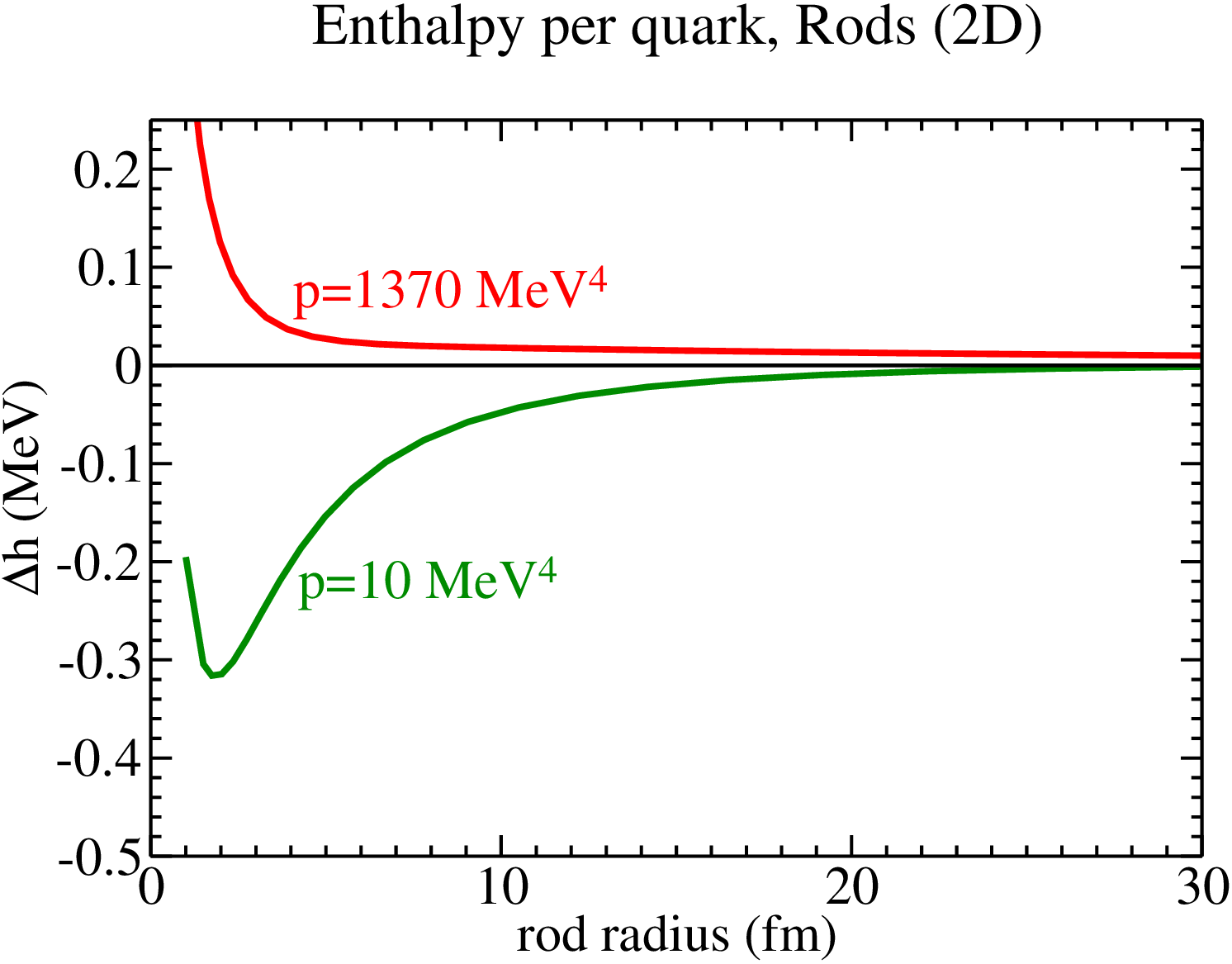}
\includegraphics[scale=0.23]{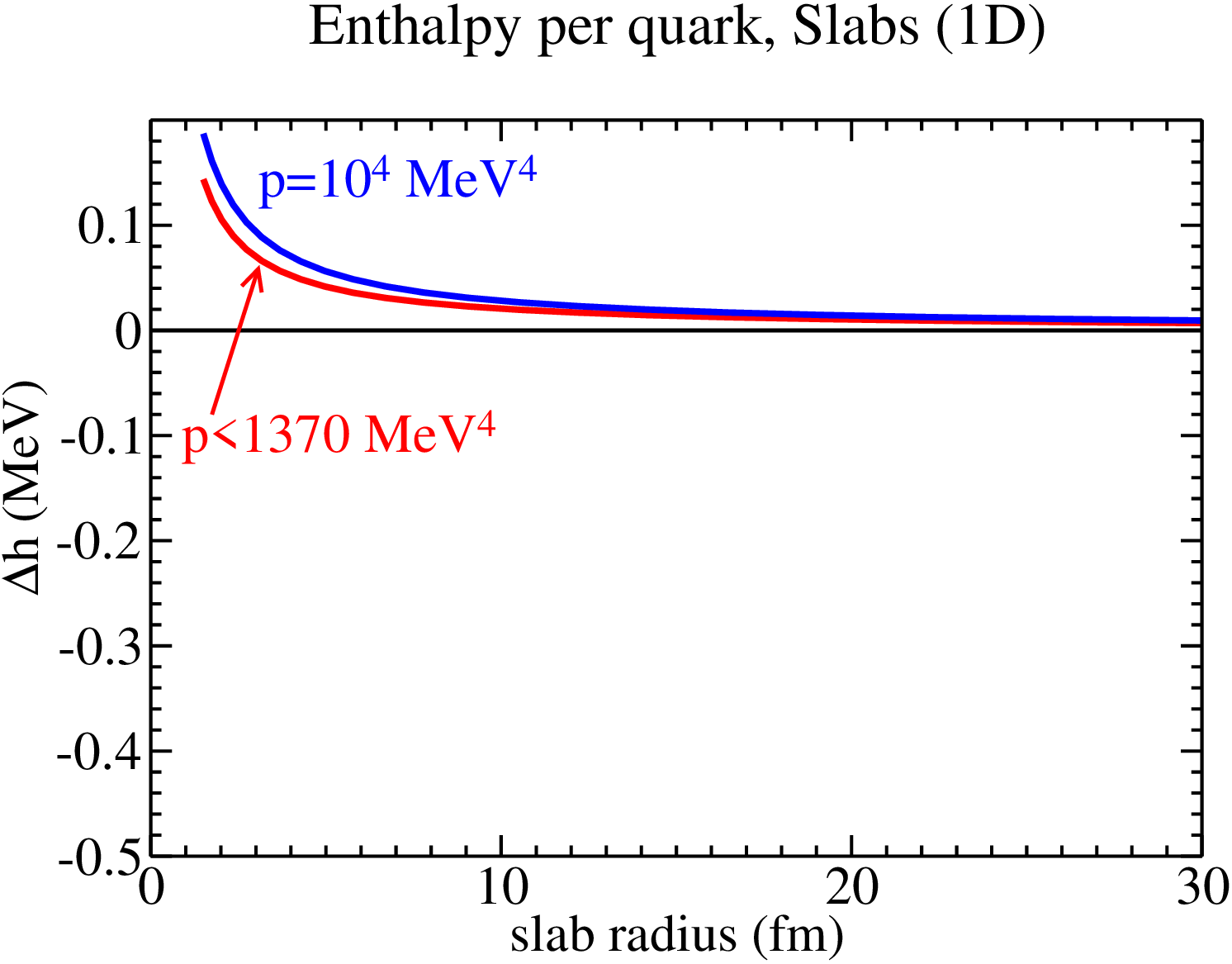}
\caption{
Search for a stable mixed phase at different pressures.
We show the excess enthalpy per quark $\De h$ \eqn{Delta_h}
as a function of the radius $R$ of the quark matter region
in the center of the cell. At each value of $R$,
the cell radius has been chosen to give
the desired pressure. The favored configuration is the one with the
smallest $\De h$, where $\De h=0$ for uniform neutral quark matter
of the same pressure.
For 3D cells, we find stable mixed phases (minima with negative $\De h$)
for pressures below a critical value, $\pcrit=1370~\MeV^4$.
2D cells always had a higher $\De h$; 1D cells were never stable.
These plots are for a quark matter EoS 
with $\mucrit=300~\MeV$, $\la_D=4.82~\fm$,
$\protect\nQ=0.0791~\fm^{-3}$, corresponding to free quarks with a
moderately heavy strange quark ($m_s=200~\MeV$ in \eqn{unpaired}). 
The surface tension was $\si=0.3~\MeV\fm^{-2}$.
}
\label{fig:stability-search}
\end{figure*}

\section{Numerical results}
\label{sec:numerical}

To get a good estimate of how thick strange star crusts might be,
we vary $\mucrit$, $\nQ$, and $\chiQ$ in the quark matter equation
of state \eqn{unpaired}, and the surface tension $\si$, over a 
physically reasonable range, calculating the crust thickness in each case.
The results of our calculations are displayed in Tables~\ref{tab:crusts}
and \ref{tab:factors}.
Before discussing them, we describe how they were obtained.

\subsection{Geometry of the mixed phase}
\label{sec:geometry}

For a given quark matter equation of state, we need to
find the maximum pressure up to which a stable mixed phase exists.
At each lower value of the pressure we must
establish the geometric configuration of the mixed phase. We can then
calculate the energy density as a function of the pressure, and obtain
the crust thickness using \eqn{dr}.

We follow the procedure described at the end of
Sec.~\ref{wigner-seitz}, varying the radius of the cell $\Rcell$ and
the radius $R$ of the quark matter region at its center, to find the
cell configuration with the lowest enthalpy per quark at each value
of the pressure.
In Fig.~\ref{fig:stability-search} we show some examples of
the search for the favored cell configuration at a given pressure.
We plot the ``excess enthalpy per quark'' 
\beq
\De h = h - h_\infty \ ,
\label{Delta_h}
\eeq
as a function of $R$, for each of the three geometries, and for 
various different pressures. Here $h$ is the enthalpy per quark
\eqn{eq:enthalpy} in a given cell, and and $h_\infty$ is its value
in uniform neutral quark matter {\em of the same pressure}. 
Configurations with negative $\De h$ therefore correspond to stable mixed
phases at the given pressure.

The results in Fig.~\ref{fig:stability-search} are for quark
matter with $\mucrit=300~\MeV$, $\nQ=0.0791$ (corresponding to $m_s=200~\MeV$ 
in \eqn{unpaired}), and $\lambda_D=4.82~\fm$ (again, a value 
appropriate to free quark matter with $\mu=300~\MeV$). The surface
tension is $0.3~\MeV/\fm^2$.
The first panel of Fig.~\ref{fig:stability-search} 
shows $\De h(R)$ for 3D cells.
The upper (red) curve is $\De h(R)$ at the critical
pressure $\pcrit=1370~\MeV^4$, which is defined by the
presence of a minimum with $\De h=0$ (at $R\approx 10~\fm$ in
this case). We also show $\De h(R)$ at
a lower pressure, $p=10~\MeV^4$; now there is a clearly favored
mixed phase, with strange droplets of radius $R\approx 4~\fm$.
If we push the pressure down to zero then the cell size goes to
infinity, and the minimum in the $\De h(R)$ curve moves further down
to around $-0.75~\MeV$.
In the second panel we show $\De h(R)$ for 2D cells at the same
two pressures. It is clear that the 2D structure has a lower
critical pressure, and at these two pressures it is energetically
unfavored relative to the 3D structure.
In the third panel we show $\De h(R)$ for 1D cells. These appear
to be even less favored. At $p=1370~\MeV^4$ the $\De h(R)$ curve is
already almost at its zero-pressure limit, which is never negative
and therefore allows no mixed phase. We had to show $\De h(R)$ for
a higher pressure, $p=10^4~\MeV^4$, to see any change in the curve.
We conclude that for this quark matter equation of state and surface tension,
and at the pressures studied in Fig.~\ref{fig:stability-search},
the only mixed phase that occurs is the 3D (droplet) one.

We note the following features of the favored configuration of
the Wigner-Seitz cell:
\begin{tightlist}{$\bullet$}
\item Increasing the pressure disfavors mixed phases: the $\De h(R)$ 
curve rises and minima are smoothed out. We hypothesize that this is because
the pressure is determined by the value of $\mue$ at the edge of the cell
\eqn{pext}; if $\mue(\Rcell)$ is increased then, because $\mue(r)$ is monotonic,
$\mue$ in the quark matter is also larger (closer to $\mue^{\rm neutral}$).
But this decreases
the energy benefit of making a mixed phase, which arises from the
departure from neutrality.
\item As the dimensionality of the mixing geometry decreases from 3 to 1,
mixed phases become less favored (at least in this range of pressures).
We hypothesize that this is because in lower dimensional
structures, a smaller proportion of the quark matter is near the
surface, where $\mue$ is different from $\mue^{\rm neutral}$.
\end{tightlist}

\begin{figure}
\includegraphics[scale=0.32]{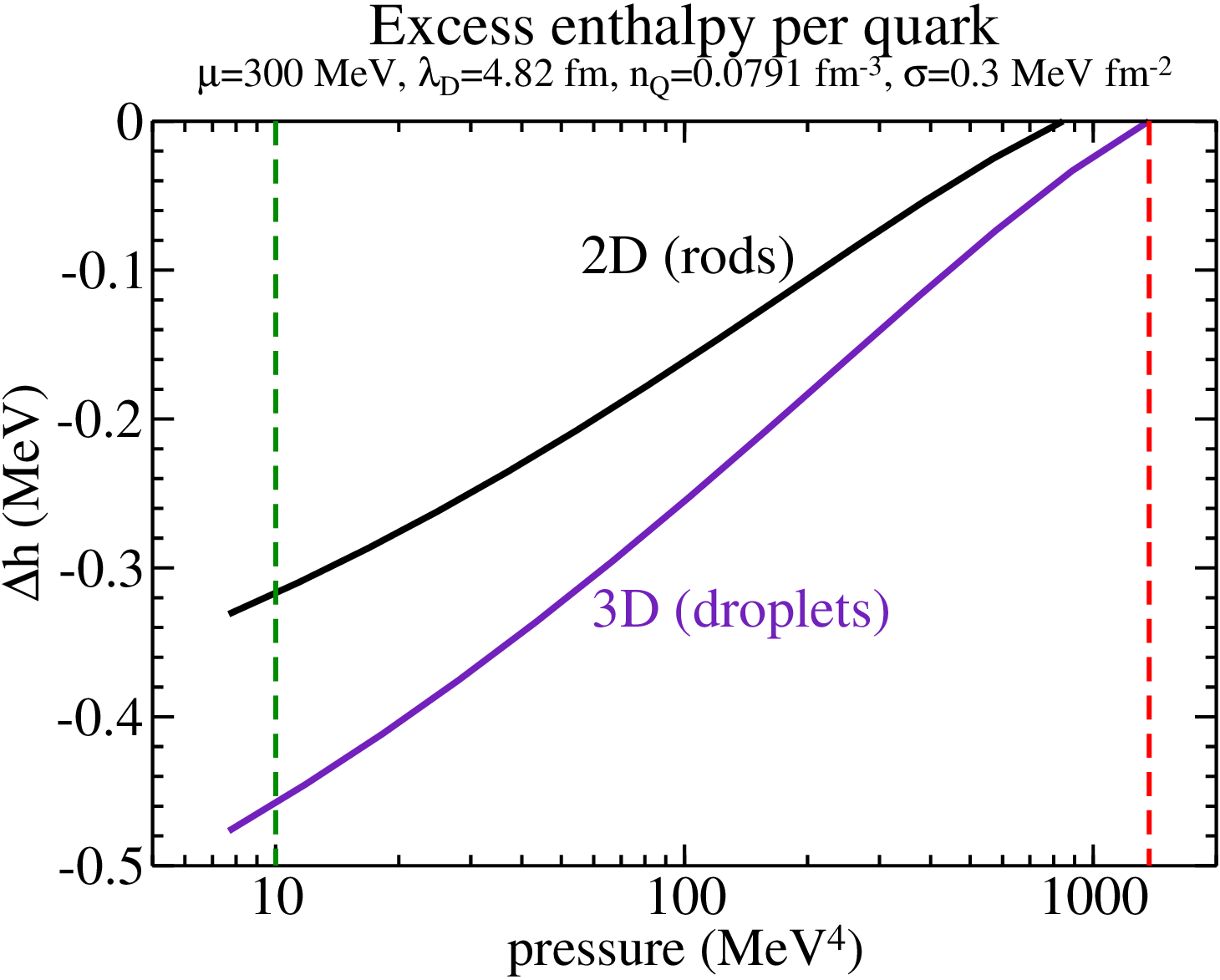}
\caption{
Minimum value of $\De h$ as a function of pressure, for quark matter
with the same characteristics as in Fig.~\ref{fig:stability-search}.
Dashed lines
mark the pressures ($10~\MeV^4$ and $1370~\MeV^4$) that were
used in Fig.~\ref{fig:stability-search}.
We see that for this type of quark matter
the 3D (droplet) structure is always energetically favored over rods.
The slab structure is never stable.
}
\label{fig:p-diffh}
\end{figure}

Since it is only the minima of $\De h(R)$ that are physically important,
we focus on them, and in Fig.~\ref{fig:p-diffh} we plot the value of
$\De h$ at the minimum as a function of pressure. The vertical dashed
lines mark the pressures $p=10~\MeV^4$ and $p=1370~\MeV^4$
used in Fig.~\ref{fig:stability-search}. We see that, for the
values of the quark matter parameters studied in
these figures, only droplet (3D) mixed phases will
occur: slabs are never stable, and rods are never favored over droplets.

\subsection{Crust thickness calculation}

In Table~\ref{tab:crusts} we show the results obtained
by repeating the calculations described above for
seven different quark matter equations of state
and four values of the surface tension. In each case,
once we have established the
energetically favored configuration of the mixed phase at each pressure,
we obtain the crust thickness using \eqn{dr}.
In order to present numbers
whose physical interpretation is clear, we assume that
the quark star has radius 10~km and mass 1.5~$M_\odot$, so we can give an
explicit crust thickness in meters. 
From the prefactor in Eq.~\eqn{dr} it is 
easy to find the multiplicative factors that
convert our values into thicknesses of
crusts on stars with different masses and radii (Table~\ref{tab:factors}). 

The first two columns of Table~\ref{tab:crusts} specify the quark matter
equation of state \eqn{eqn:generic_EoS}, by giving the value of
$\nQ$ and the value of $\la_D$ (which fixes $\chiQ$ via \eqn{Debye_length}).
We fix $\mucrit=300~\MeV$.
The third column gives the maximum surface tension $\sicrit$
for which a crust of droplets of strange matter could occur,
the fourth column gives an estimated upper limit on the thickness, and
the last four columns give the thickness
of the crust for various values of the surface tension $\si$.

\begin{table*}
\setlength{\tabcolsep}{1em}
\begin{tabular}{ccccccccc}
$\lambda_D$ & $n_Q$ & $\sicrit$(3D) & $\De R_{\rm max}$ 
& \multicolumn{4}{c}{$\De R$ (m) at}  \\
   (fm)     & (fm${}^{-3}$) & (MeV fm${}^{-2}$) & (m) & $\si\!=\!0.3$ &  $\si\!=\!1.0$ 
  & $\si\!=\!3.0$ &  $\si\!=\!10.0$ &  \\[1ex]
4.82  & 0.0445 & 0.533 & 36 & 9  & -- & -- & --  \\ 
4.82  & 0.0791 & 1.69  & 120 & 67 & 25 & -- & --  \\
4.82  & 0.124 & 4.12  & 280 & 220  & 160 & 39 & --  \\[1ex]
6.82  & 0.0445 & 1.51 & 72 & 40  &  13  &  -- & -- \\
6.82  & 0.0791 & 4.8  & 230 & $>$\,170  & 120 & 45 & -- \\
6.82  & 0.124 & 11.6  & 550 & $>$\,460  & $>$\,390 & 280 & 39 \\[1ex]
9.65  & 0.0445 & 4.27 & 140 & 110  & 75 & 22 & --  \\
\end{tabular}
\caption{Crust thickness $\De R$ (in meters) for a strange star of radius 10~km
and mass 1.5~$M_\odot$; for more general stars see Table~\ref{tab:factors}.
We calculate the crust for seven different quark matter equations of state
and four values of the surface tension.
The first two columns, $\la_D$ and $\protect\nQ$,
specify the quark matter equation of state \eqn{eqn:generic_EoS}
(via \eqn{Debye_length}). The third column gives the maximum surface tension
for which a strangelet crust will occur \eqn{sicrit_result}.
The fourth column gives a rough upper limit on $\De R$ using an estimate from
Ref.~\cite{Jaikumar:2005ne}.
The last four columns give our results for the crust thickness
in meters for different values
of the surface tension  $\si$ (given in MeV\,fm${}^{-2}$)
of the interface between quark matter and vacuum.
The crust thickness is very sensitive to the
equation of state and the surface tension. It ranges from zero
up to several hundred meters.
} 
\label{tab:crusts}
\end{table*}

\begin{table}
\setlength{\tabcolsep}{1em}
\begin{tabular}{ccc}
R (km) & M ($M_\odot$) & $\De R/(\De R\ \mbox{in Table~\ref{tab:crusts}})$ \\
 4 & 0.05 & 8.3 \\
 6 & 0.2 & 4.4 \\
 8 & 0.5 & 2.8 \\
10 & 1.0 & 2.0
\end{tabular}
\caption{Correction factors to be applied to crust thickness values in
Table~\ref{tab:crusts}, for quark stars of various radii and
masses.  These follow from the factor multiplying the integral
in Eq.~\eqn{dr}. Note that smaller stars have thicker crusts.
}
\label{tab:factors}
\end{table}

\subsubsection{Range of parameters studied}

Our assumption that the strange matter hypothesis is valid
requires that $\mucrit$ must be less than
the quark chemical potential of nuclear matter, about 310~MeV,  so we
fix $\mucrit=300~\MeV$. The value of $\mu$ inside our strange matter lumps
will always be within a few MeV of $\mucrit$, because in order to
get any crust at all the surface tension
cannot be large enough to cause significant compression.

Typical values of $\chiQ$ will be
around $0.2 \mucrit^2$ \eqn{unpaired},
corresponding to $\la_D\approx 4.82~\fm$.
Ref.~\cite{Alford:2006bx} found that in the 2SC phase
$\chiQ$ is smaller by a factor of 2. In Table~\ref{tab:crusts}
we explore this range, using three values, $\chiQ=0.2 \mucrit^2$,
$\chiQ=0.1 \mucrit^2$,
and $\chiQ=0.05 \mucrit^2$, corresponding to $\la_D=4.82~\fm$,
$\la_D=6.82~\fm$, and and  $\la_D=9.65~\fm$ (via \eqn{Debye_length}).

Typical values of $\nQ$ will be around  $0.05\mucrit m_s^2$
\eqn{unpaired}, and a reasonable range would correspond to varying
$m_s$ over its physically plausible range, from about 100 to 300 MeV.
(To have strange matter in the star, $m_s$ must be less than $\mucrit$.)
In Table~\ref{tab:crusts} we use $\nQ=0.0445$, $0.0791$, and
$0.124~\fm^{-3}$,
which would correspond to $m_s=150$, 200, and 250~MeV in
\eqn{unpaired}.
For $\la_D=9.65~\fm$ we only show results for  $\nQ=0.00445~\fm^{-3}$.
We do not show results for
$\nQ=0.0791~\fm^{-3}$ or $\nQ=0.124~\fm^{-3}$, because in those cases
the value of $\mue^{\rm neutral}$ would be 133~MeV and 208~MeV respectively,
which violates our assumption that
$\mue\ll\mu$, and is also 
above the muon mass $m_\mu=105.66~\MeV$,
so we would have to take into account muons as well as
electrons. 


The value of the maximum surface tension $\sicrit$
for which a crust of droplets of strange matter could occur
(third column of Table~\ref{tab:crusts})
follows from \eqn{sicrit_result}. It is the maximum 
surface tension at which
an isolated (zero-pressure) droplet would have lower enthalpy per quark
than neutral quark matter at zero pressure,
i.e.~at the onset phase transition at $\mu=\mucrit$.
The last four columns of Table~\ref{tab:crusts} give our results for
the thickness
of the crust at a range of values of the surface tension $\si$.
For values of $\si$ above $\sicrit$ there is no crust.
The values of $\si$ that we use are physically reasonable,
given that rough estimates of surface tension from the
bag model are in the range $4$ to $10~\MeV/\fm^2$ 
\cite{Berger:1986ps,PhysRevC.44.566.2}.

\begin{figure}
\includegraphics[scale=0.32]{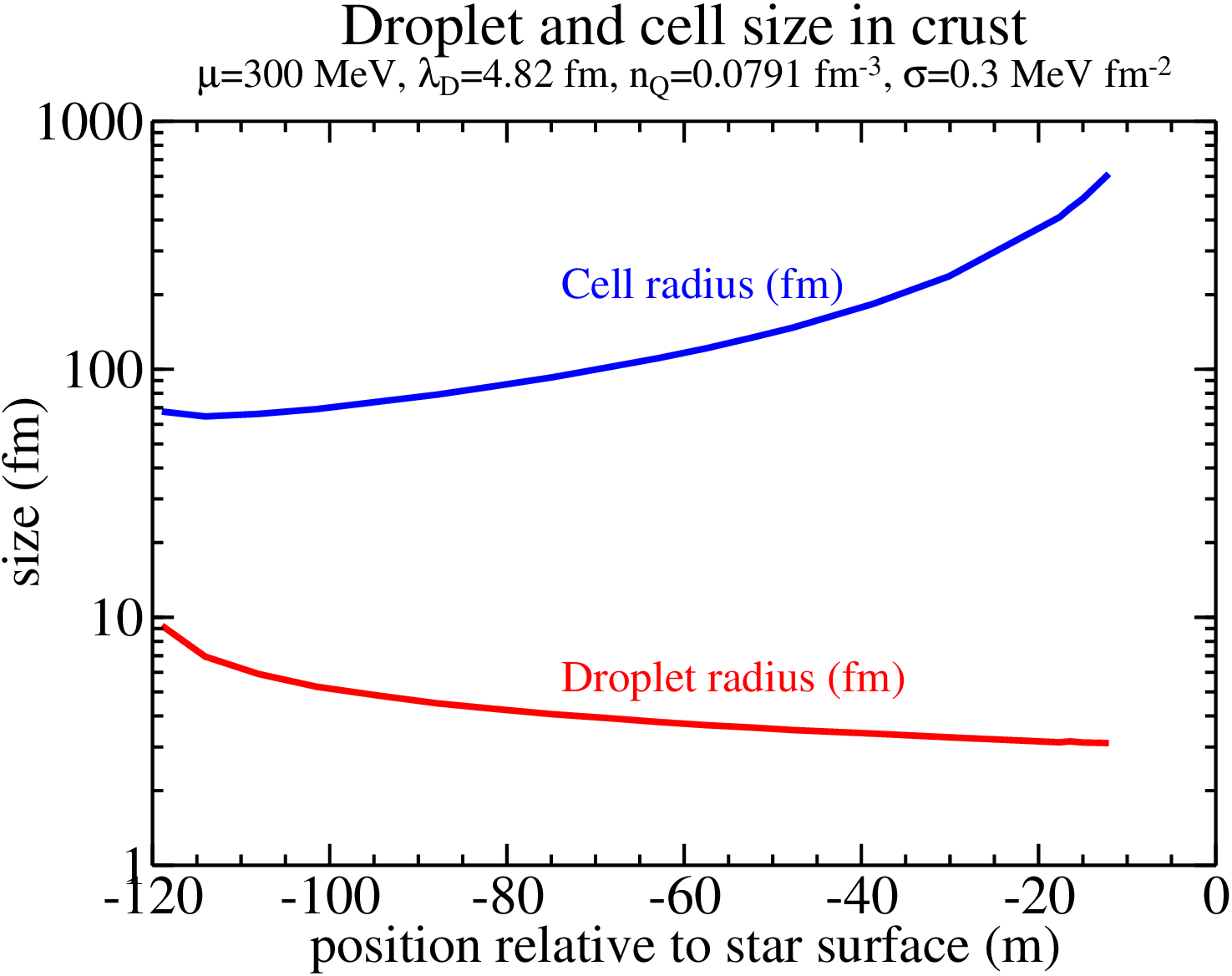}
\caption{
Size of the quark matter droplets and
the Wigner-Seitz cell (in fm) as a function of position
in the the 120m-thick crust that we predict for quark matter with
the stated parameter values, which correspond to free quark matter
with a moderately heavy strange quark ($m_s=200~\MeV$ in
\eqn{unpaired}).
}
\label{fig:crust}
\end{figure}

\subsubsection{Crust thickness results}
\label{sec:thickness-results}

In Table~\ref{tab:crusts}, we present the results of our calculation
of the crust thickness as a function of the quark matter parameters.
Clearly the thickness is very sensitive to the values of these parameters.

In Table~\ref{tab:factors}
we give the correction factors for some typical masses and radii 
that are expected for quark stars (see, for example, 
Ref.~\cite{Weber:2004kj}, Fig.~28). We see that smaller and lighter
quark stars have thicker crusts. In fact, for a star of radius
4~km \cite{Fraga:2001id}
the crust thickness could easily be of order 1000~m, at which point
the assumption $\De R/\Rstar$ used in Eq.~\eqn{dr} begins to become
questionable. If the crust is thick enough then one must keep
higher powers of $\De R/\Rstar$ in calculating it, and it may even be
necessary to solve the
full Tolman Oppenheimer Volkoff equation 
\cite{Tolman:1939jz,Oppenheimer:1939ne}.

Some of the crust thicknesses in Table~\ref{tab:crusts}
are given as lower limits. This is because for large $\la_D$ and $\nQ$,
and low surface tension, mixed phases are highly favored, and persist
up to extremely high pressures (of order $10^6~\MeV^4$). To achieve such
a high pressure requires a very ``cramped'' cell geometry, in which
$\Rcell$ is only a little larger than the quark droplet radius $R$.
In such a geometry we can no longer trust our
approximation of treating the cell as spherical,
rather than a unit cell of a crystal lattice (e.g.~cubic).
We therefore use an alternative
upper limit $p_{\rm max}$ on the pressure: $p_{\rm max}$ is the pressure
below which the favored cell configuration always has $\Rcell>2R$.
We can then calculate a lower limit on the crust thickness, by integrating
\eqn{dr} up to $p_{\rm max}$.

In the limit of low surface tension, our results are compatible with
the upper limit $\De R_{\rm max}$ (fourth column in Table~\ref{tab:crusts})
which follows from taking the estimate
obtained by ignoring surface tension and Debye screening
in Ref.~\cite{Jaikumar:2005ne} (their Eq.~(13)) and applying
the TOV correction factor $(1-2GM/\Rstar)$.
We find that even a relatively moderate surface tension, around 1 to
$10~\MeV\fm^{-2}$, reduces the thickness of the crust or even
eliminates it completely.  The mechanism is clear: larger values of
the surface tension disfavor the mixed phase by increasing surface
energy costs, leading to a lower $\pcrit$ and thinner crusts.
However, it remains possible that a quark star could have a crust
hundreds of meters thick.

We find that the crust is thickest for large values of $\nQ$ and
$\la_D$, as one would expect from
the estimate $\De R_{\rm max}\propto n_Q^2/\chiQ 
\propto n_Q^2\la_D^2$ \cite{Jaikumar:2005ne}. This is consistent
with the observation from
Fig.~\ref{fig:stability-search} that we can measure how favored
a mixed phase is by the depth $\De h_{\rm min}$ of the minimum in $\De h(R)$.
Since the $\De h(R)$ curve moves up as the pressure increases (see 
Sec.~\ref{sec:geometry}), we can guess that the deeper the minimum
at zero pressure, the thicker the crust. From Ref.~\cite{Alford:2006bx}
(Fig.~2 and Eq.~(26)), we find that at a given $\si$, the depth of
the minimum is $\De h_{\rm min}\propto -n_Q^2/\chiQ$, 
i.e.~$\De h_{\rm min}\propto -n_Q^2\la_D^2$. Hence
the mixed phase is more favored, and the crust is thicker,
for larger values of $n_Q$ and $\la_D$.

All the crusts in Table~\ref{tab:crusts} consist entirely of
3D structures, i.e.~spherical droplets of quark matter in a
neutralizing background of electrons. We never found any pressure
for any quark matter parameters where 2D (rod) or 1D (slab) structures
were energetically preferred. It would be interesting to see whether
this remains true when the cell is allowed to be a different shape
from the strangelet (e.g.~square or cubic cells).

\subsubsection{Internal structure of the crust }
\label{sec:crust-structure}

In Fig.~\ref{fig:crust} we select one of our quark matter equations of
state, and
show how the properties of the strangelet crystal lattice vary with
position in the crust. The horizontal axis is $\De r = r-\Rstar$,
so more negative values correspond to deeper parts of the crust.
The plot should end at $\De r=0$, but we were not able to push our
numerical calculations to that value, so the curves end at 
$\De r\approx -10$~m. 

As one approaches the surface, the size  $\Rcell$ of the Wigner-Seitz cell
grows very large (note that Fig.~\ref{fig:crust} uses a logarithmic scale
for the vertical axis). 
This means that the strangelet density becomes very low. We expect that as
$\De r\to 0$, $\Rcell$ will diverge, since the pressure must go to zero,
so $\mue$ at the edge of a cell must go to zero, so the cell size must 
become infinite. In this limit the droplets in the crust
effectively become isolated strangelets, and we expect their size to
settle down to that of the most stable isolated strangelet for this 
form of quark matter. We can predict this value
from Ref.~\cite{Alford:2006bx}, eqn (24): by minimizing the free
energy per quark $\De g/n$ (which is equivalent to $\De h$ 
in this paper) for the values of $\mucrit$, $\nQ$, $\chiQ$, and $\si$
used in Fig.~\ref{fig:crust}, we find that the most stable strangelet
has a radius of 3.0~\fm, which is exactly the asymptotic value
emerging in Fig.~\ref{fig:crust}.
For such small strangelets we expect that our Thomas-Fermi approach
is not accurate, and shell model corrections will become important:
including such corrections is a topic for future research.

As one goes deeper into the star ($\De r$ becoming more negative), the
pressure rises, so the cell size decreases, and the droplet size
slowly increases, until we reach the critical pressure at which
uniform neutral quark matter becomes favored over the strangelet
lattice. This is a first-order phase transition, as is clear from
Fig.~\ref{fig:stability-search}, so the curves in Fig.~\ref{fig:crust}
end suddenly, without any singular behavior.  In this paper we do not
take in to account the possibility of a metastable lattice that might
persist to higher pressures.

Fig.~\ref{fig:crust} shows that the strangelet crystal crust of a quark
star tends to be fairly dilute: over most of the crust the quark matter
droplet size is small, of order the Debye length in quark matter $\la_D$,
while the cell size is larger, by a factor of 10 or more.

\section{Discussion}
\label{sec:discussion} 

The calculations described in this paper give us a more precise
picture of the strangelet-crystal crust of a quark star. 
The results presented in Table~\ref{tab:crusts} show that the thickness
of the strangelet-crystal crust of a strange star is very sensitive to the
surface tension $\si$ of the interface between quark matter and the vacuum,
and to the quark matter parameters  $\nQ$ and $\chiQ$ \eqn{eqn:generic_EoS}, 
which determine
the response of the quark matter to deviations of the
electrostatic potential from its neutrality value.
Our results are compatible with those of Ref.~\cite{Jaikumar:2005ne}
where an upper limit on the crust thickness was obtained by
ignoring surface tension and Debye screening.
The crust is thickest for large
$\nQ$ and small $\chiQ$ (large $\la_D$).
As discussed in Sec.~\ref{sec:thickness-results}, we find that
values of surface tension in the physically expected range, around 1 to
$10~\MeV\fm^{-2}$, reduce the thickness of the crust and may
even eliminate it completely, but it remains possible that
a quark star of radius 10~km
could have a crust several hundred meters thick. From
Table~\ref{tab:factors} we see that for a smaller star the crust could
be even thicker. 

The geometry of the mixed phase in our crusts, on the other hand, shows
no variation at all. It is always three-dimensional, containing
spherical droplets
of quark matter. We never find any case where a two-dimensional (rod)
or one-dimensional (slab) geometry is favored.

Our calculations and results suggest two directions for future work:
firstly, one could study phenomenological consequences of our
understanding of the strangelet-crystal crust of a quark star.
Secondly, one could improve on our treatment of the strangelet crystal,
by relaxing some of the assumptions listed in Sec.~\ref{sec:intro}.

The most obvious phenomenological task is to
revisit computations of the frequencies of
seismic vibrations which are observed after giant flares in magnetars
\cite{Watts:2006hk,Chugunov:2006kk}. Ref.~\cite{Watts:2006hk} found
that the strangelet crystal crust did not have the right spectrum of
toroidal shear modes to account for current observations: it would be
interesting to see whether taking in to account the surface tension
and Debye screening affects that conclusion. 
Other aspects of the phenomenology of the crust could
also be studied, for example (a) the thermal response
of the crust to accretion \cite{2000ApJ...531..988B};
(b) the role of the crust in the trapping of neutrinos and photons just after
a type-II supernova \cite{Burrows:2004vq};
(c) the spectrum of
photons radiated from the surface of a quark star
\cite{Page:2002bj,Jaikumar:2004zy,Harko:2004ts};
(d) the contribution of the crust to the moment of inertia and glitches
\cite{1992ApJ...400..647G}; 
(e) the damping of $r$-modes in by
shear viscosity in the crust \cite{2000ApJ...529L..33B,Caballero:2008tx} (for
quark stars, the contribution from the interior has been calculated 
\cite{Madsen:1999ci,Jaikumar:2008kh});
(f) the thermal relaxation time of the crust and its response to the
post-supernova ``cooling wave'' \cite{Gnedin:2000me}. 
The thermal relaxation time of the crust depends on the thermal conductivity, 
for which
we can make a very rough estimate using appendix A of 
Ref.~\cite{Gnedin:2000me}. We find values of order a few hundred $\MeV^2$ at
$T\sim 0.1~\MeV$, which is comparable to the range 
$10^{18}~{\rm erg}\,{\rm cm}^{-1}{\rm s}^{-1}{\rm K}^{-1}$ for 
low-density nuclear matter
(Ref.~\cite{Gnedin:2000me}, Fig.~4). We defer
a more accurate calculation to future work.

To improve on our treatment, the most pressing issues are to
use a realistic shape for the Wigner-Seitz cells (which should be
unit cells of some regular lattice, rather than spheres), to
include shell-model corrections for the smallest strangelets, and to
allow for ``Swiss-cheese'' phases where most of the unit cell
consists of quark matter, with a hole at the center containing
electrons. 

As discussed in Sec.~\ref{sec:thickness-results}, the
shape of the cell becomes important at very high pressures, and
our use of the spherical approximation meant that in some
cases we could only obtain lower limits on the crust thickness.
Studying more realistic shapes is straightforward in principle,
but would require a more demanding
multidimensional numerical solution of the Poisson equation.

Shell-model corrections can be of order one MeV per quark for strangelets
of size $R\lesssim 5~\fm$ \cite{Madsen:1994vp,Amore:2001uf},
which is not negligible relative to
our typical enthalpy per quark (Fig.~\ref{fig:stability-search}),
and may therefore affect our results for the outer part of the
crust, where we predict strangelets as small as 3~fm (Fig.~\ref{fig:crust}).

Treating Swiss-cheese phases would require us to 
include curvature energy
as well as surface tension. This highlights the fact that
we treated the
interface between quark matter and vacuum in a very simplified way,
as a zero-width interface with a surface tension. However, since the
quark confinement distance is about 1~fm, the interface might well
have structure on this distance scale. Like the shell-model
effects described above, this could be relevant to
the low-pressure regime, where the strangelets
can be as small as a few fm.
There are even indications that when such physics is taken
into account, the CFL phase may undergo some degree of charge separation
\cite{Madsen:2001fu,Oertel:2008wr}, 
raising the possibility that there might be some
sort of crystalline crust on quark stars made of CFL quark matter.

\section*{Acknowledgments}

We thank Junhua Chen, Prashanth Jaikumar,
Jes Madsen, Sanjay Reddy for useful discussions. 
This research was
supported in part by the Offices of Nuclear Physics and High
Energy Physics of the Office of
Science of the U.S.~Department of Energy under contracts
\#DE-FG02-91ER40628,  
\#DE-FG02-05ER41375, 
and by the Undergraduate Research
Office and McDonnell Center for Space
Sciences of Washington University in St. Louis.

\bibliographystyle{h-physrev4}
\bibliography{crust} 

\begin{thebibliography}{10}

\bibitem{Bodmer:1971we}
A.~R. Bodmer,
\newblock Phys. Rev. {\bf D4}, 1601 (1971).

\bibitem{Witten:1984rs}
E.~Witten,
\newblock Phys. Rev. {\bf D30}, 272 (1984).

\bibitem{Weber:2004kj}
F.~Weber,
\newblock Prog. Part. Nucl. Phys. {\bf 54}, 193 (2005), [astro-ph/0407155].

\bibitem{Farhi:1984qu}
E.~Farhi and R.~L. Jaffe,
\newblock Phys. Rev. {\bf D30}, 2379 (1984).

\bibitem{Alcock:1986hz}
C.~Alcock, E.~Farhi and A.~Olinto,
\newblock Astrophys. J. {\bf 310}, 261 (1986).

\bibitem{Stejner:2005mw}
M.~Stejner and J.~Madsen,
\newblock Phys. Rev. {\bf D72}, 123005 (2005), [astro-ph/0512144].

\bibitem{Usov:1997eg}
V.~V. Usov,
\newblock Astrophys. J. {\bf 481}, L107 (1997), [astro-ph/9703037].

\bibitem{Jaikumar:2005ne}
P.~Jaikumar, S.~Reddy and A.~W. Steiner,
\newblock Phys. Rev. Lett. {\bf 96}, 041101 (2006), [nucl-th/0507055].

\bibitem{Alford:2006bx}
M.~G. Alford, K.~Rajagopal, S.~Reddy and A.~W. Steiner,
\newblock Phys. Rev. {\bf D73}, 114016 (2006), [hep-ph/0604134].

\bibitem{Alford:2000ze}
M.~G. Alford, J.~A. Bowers and K.~Rajagopal,
\newblock Phys. Rev. {\bf D63}, 074016 (2001), [hep-ph/0008208].

\bibitem{Casalbuoni:2003wh}
R.~Casalbuoni and G.~Nardulli,
\newblock Rev. Mod. Phys. {\bf 76}, 263 (2004), [hep-ph/0305069].

\bibitem{Ravenhall:1983uh}
D.~G. Ravenhall, C.~J. Pethick and J.~R. Wilson,
\newblock Phys. Rev. Lett. {\bf 50}, 2066 (1983).

\bibitem{Alford:2007xm}
M.~G. Alford, A.~Schmitt, K.~Rajagopal and T.~Schafer,
\newblock 0709.4635.

\bibitem{Heiselberg:1993dc}
H.~Heiselberg,
\newblock Phys. Rev. {\bf D48}, 1418 (1993).

\bibitem{Glendenning:1992vb}
N.~K. Glendenning,
\newblock Phys. Rev. {\bf D46}, 1274 (1992).

\bibitem{Alford:2004hz}
M.~Alford, C.~Kouvaris and K.~Rajagopal,
\newblock Phys. Rev. {\bf D71}, 054009 (2005), [hep-ph/0406137].

\bibitem{Tolman:1939jz}
R.~C. Tolman,
\newblock Phys. Rev. {\bf 55}, 364 (1939).

\bibitem{Oppenheimer:1939ne}
J.~R. Oppenheimer and G.~M. Volkoff,
\newblock Phys. Rev. {\bf 55}, 374 (1939).

\bibitem{Maruyama:2007ey}
T.~Maruyama, S.~Chiba, H.-J. Schulze and T.~Tatsumi,
\newblock Phys. Rev. {\bf D76}, 123015 (2007), [0708.3277].

\bibitem{Madsen:1994vp}
J.~Madsen,
\newblock Phys. Rev. {\bf D50}, 3328 (1994), [hep-ph/9407314].

\bibitem{Amore:2001uf}
P.~Amore, M.~C. Birse, J.~A. McGovern and N.~R. Walet,
\newblock Phys. Rev. {\bf D65}, 074005 (2002), [hep-ph/0110267].

\bibitem{Madsen:2000kb}
J.~Madsen,
\newblock Phys. Rev. Lett. {\bf 85}, 4687 (2000), [hep-ph/0008217].

\bibitem{Madsen:2001fu}
J.~Madsen,
\newblock Phys. Rev. Lett. {\bf 87}, 172003 (2001), [hep-ph/0108036].

\bibitem{Madsen:2008bx}
J.~Madsen,
\newblock Phys. Rev. Lett. {\bf 100}, 151102 (2008), [0804.2140].

\bibitem{Oertel:2008wr}
M.~Oertel and M.~Urban,
\newblock Phys. Rev. {\bf D77}, 074015 (2008), [0801.2313].

\bibitem{Christiansen:1997rc}
M.~B. Christiansen and N.~K. Glendenning,
\newblock Phys. Rev. {\bf C56}, 2858 (1997), [astro-ph/9706056].

\bibitem{Christiansen:1997vt}
M.~B. Christiansen and J.~Madsen,
\newblock J. Phys. {\bf G23}, 2039 (1997).

\bibitem{Lyubarsky:2002cs}
Y.~Lyubarsky, D.~Eichler and C.~Thompson,
\newblock Astrophys. J. {\bf 580}, L69 (2002), [astro-ph/0211110].

\bibitem{Weisstein}
E.~W. Weisstein,
\newblock Jacobi elliptic functions.,
\newblock From MathWorld--A Wolfram Web Resource.
  \url{http://mathworld.wolfram.com/JacobiEllipticFunctions.html}.

\bibitem{Berger:1986ps}
M.~S. Berger and R.~L. Jaffe,
\newblock Phys. Rev. {\bf C35}, 213 (1987).

\bibitem{PhysRevC.44.566.2}
M.~S. Berger and R.~L. Jaffe,
\newblock Phys. Rev. C {\bf 44}, 566 (1991).

\bibitem{Fraga:2001id}
E.~S. Fraga, R.~D. Pisarski and J.~Schaffner-Bielich,
\newblock Phys. Rev. {\bf D63}, 121702 (2001), [hep-ph/0101143].

\bibitem{Watts:2006hk}
A.~L. Watts and S.~Reddy,
\newblock Mon. Not. Roy. Astron. Soc. {\bf 379}, L63 (2007),
  [astro-ph/0609364].

\bibitem{Chugunov:2006kk}
A.~I. Chugunov,
\newblock Mon. Not. Roy. Astron. Soc. {\bf 371}, 363 (2006),
  [astro-ph/0606310].

\bibitem{2000ApJ...531..988B}
E.~F. {Brown},
\newblock \apj {\bf 531}, 988 (2000), [arXiv:astro-ph/9910215].

\bibitem{Burrows:2004vq}
A.~Burrows, S.~Reddy and T.~A. Thompson,
\newblock Nucl. Phys. {\bf A777}, 356 (2006), [astro-ph/0404432].

\bibitem{Page:2002bj}
D.~Page and V.~V. Usov,
\newblock Phys. Rev. Lett. {\bf 89}, 131101 (2002), [astro-ph/0204275].

\bibitem{Jaikumar:2004zy}
P.~Jaikumar, C.~Gale, D.~Page and M.~Prakash,
\newblock Int. J. Mod. Phys. {\bf A19}, 5335 (2004), [astro-ph/0407091].

\bibitem{Harko:2004ts}
T.~Harko and K.~S. Cheng,
\newblock Astrophys. J. {\bf 622}, 1033 (2005), [astro-ph/0412280].

\bibitem{1992ApJ...400..647G}
N.~K. {Glendenning} and F.~{Weber},
\newblock \apj {\bf 400}, 647 (1992).

\bibitem{2000ApJ...529L..33B}
L.~{Bildsten} and G.~{Ushomirsky},
\newblock \apjl {\bf 529}, L33 (2000), [arXiv:astro-ph/9911155].

\bibitem{Caballero:2008tx}
O.~L. Caballero, S.~Postnikov, C.~J. Horowitz and M.~Prakash,
\newblock 0807.4353.

\bibitem{Madsen:1999ci}
J.~Madsen,
\newblock Phys. Rev. Lett. {\bf 85}, 10 (2000), [astro-ph/9912418].

\bibitem{Jaikumar:2008kh}
P.~Jaikumar, G.~Rupak and A.~W. Steiner,
\newblock 0806.1005.

\bibitem{Gnedin:2000me}
O.~Y. Gnedin, D.~G. Yakovlev and A.~Y. Potekhin,
\newblock Mon. Not. Roy. Astron. Soc. {\bf 324}, 725 (2001),
  [astro-ph/0012306].

\end{thebibliography}



\end{document}